# A Review on Serious Games in E-learning

Huansheng Ning, Hang Wang, Wenxi Wang, Xiaozhen Ye, Jianguo Ding, and Per Backlund

Abstract:
E-learning is a widely used learning method, but with the development of society, traditional E-learning method has exposed some shortcomings, such as the boring way of teaching, so that it is difficult to increase the enthusiasm of students and raise their attention in class. The application of serious games in E-learning can make up for these shortcomings and effectively improve the quality of teaching. When applying serious games to E-learning, there are two main considerations: educational goals and game design. A successful serious game should organically combine the two aspects and balance the educational and entertaining nature of serious games. This paper mainly discusses the role of serious games in E-learning, various elements of game design, the classification of the educational goals of serious games and the relationship between educational goals and game design. In addition, we try to classify serious games and match educational goals with game types to provide guidance and assistance in the design of serious games. This paper also summarizes some shortcomings that serious games may have in the application of E-learning.

# 1. Introduction

Nowadays, learning resources on the Internet are becoming more and more abundant. E-learning is a learning method that is widely used all over the world. During the raging period of COVID-19, E-learning ensured the completion of school curriculums and achieved remarkable results. Traditional E-learning usually includes four elements: 1) teaching content through live broadcast or multimedia courseware prepared in advance; 2) course tests available on the website; 3) the function of correcting homework; 4) the function of communication and discussion.

However, with the rapid innovation of technology, education field is also developing towards a diversified, personalized and entertaining learning. The problems of the traditional E-learning method are gradually exposed and great challenges arose. The traditional E-learning method is faced with the following challenges:

1) Teaching methods are difficult to increase students' enthusiasm for learning. In the traditional E-learning method, the educational content mainly relies on multimedia files to convey, and this teaching method can easily make students feel bored.

2) Unable to supervise students' learning process, and the teaching effect is heavily depends on the students' self-learning ability. To guarantee effective e-learning, students must have good self-management and self-control skills. For students who have not developed good study habits and learning attitudes, the teaching effect of traditional E-learning is often poor.

3) Low participation and emotional alienation. Traditional e-learning methods lack interaction, cannot provide a good teaching atmosphere, and cannot allow teachers and students to get timely feedback.

The application of serious games in education field provides a good solution to the problems faced by the traditional E-learning method. At present, serious games that are not purely used for entertainment [1] have been widely used in lots of fields such as disaster relief, medical treatment and education [2]. In recent years, the term "serious games" has been dominant in the field of game research [3]. As one of the popular technologies, the seriousness (educational) and entertaining nature of serious games make it an effective tool for education and teaching. Serious games have the characteristics of relaxation and entertaining [4], and at the same time impart knowledge and skills to players. In fact, serious games can produce better teaching effects than other methods [5], and they play a unique role in education.

The application of serious games in E-learning makes up for the shortcomings of traditional E-learning methods. Serious games can play the following roles in the field of E-learning: 1) Serious games can effectively increase students' learning enthusiasm. Research shows that deep learning can be achieved by solving the emotional and intrinsic motivation of students participating in E-learning. Serious games can attract students to complete their teaching goals through fun. And the game mechanics of serious games (such as competition mechanism, reward and punishment mechanism, etc.) can also effectively motivate students. 2) The way of delivering educational content through serious games can effectively improve teaching efficiency and teaching quality. 3) Serious games can provide good interactivity and experience. Therefore, serious games have the potential to become a good and important teaching tool.

By definition, the main purpose of serious games is to impart knowledge and skills. And when serious games are applied to the field of E-learning, more attention should be paid to the educating and teaching effects of serious games. Serious games should not only help increase students' enthusiasm for learning, but also optimize students' learning methods to achieve effective learning. Therefore, the design of serious games needs the guidance of pedagogic philosophy, and the development of serious games should be based on educational goals. And in terms of design and development, each task in a serious game must focus on a specific learning goal and establish a balance between serious and interesting content [3].

The key to combining serious games and E-learning is to achieve a balance and close connection between serious aspect (educational aspect) and entertaining aspect.

However, in the application of many current serious games in E-learning, the design of serious games often does not pay enough attention to education, or does not fully consider how to establish a balance between serious and game aspects, so that they cannot achieve desirable teaching effect. In response to this problem, this paper analyzes and summarizes the design and development of serious games in two aspects: game aspect and education aspect. And in these two parts, the main content that needs to be considered in game aspect is game design, and the main content that needs to be considered in education is educational goals. Therefore, this paper will introduce the game design and educational goals of serious games.

The following is the main content of this paper:

First, this paper introduces the role of serious games in E-learning and analyzes the various elements involved in serious games in game design. Based on this, we classify the serious games used in E-learning. Second, this paper introduces the classification method of educational goals (Bloom's Taxonomy), which provides a general analysis and classification method for different educational content. Third, we analyze and match the types of serious games we have summarized with the categories of educational goals to provide effective help and guidance for the development of serious games.

In Section 2, we introduce the role of serious games in E-learning, the elements involved in game design, and the classification theory of educational goals (that is, Bloom's Taxonomy, including the original and revised version).
In Section 3, this paper classifies serious games. Moreover, we also analyze and match the types of serious games with the categories of educational goals. In Section 4, this paper discusses issues that need to be considered when evaluating serious games. In Section 5, we summarize the whole paper.

# 2. Serious Games in E-learning

## 2.1 The Role of Serious Games in E-learning

The important purpose of applying serious games to E-learning is to create an interesting, challenging and participatory learning environment to convey serious content. As a teaching tool, serious games should effectively enhance students' motivation to learn, encourage learners to actively participate in the educational process [5], and persuade players to continue playing games through game mechanisms such as rewards and prompts. Studies have shown that if the elements of game design are sufficiently attractive to players, a substance called endorphins will be released into the player's body, and the player can easily complete the learning task [6]. Serious games can also establish good communication and interaction with players to increase students' participation in the learning process [7]. The teaching method of serious games can help students understand and accept new materials, improve teaching efficiency and teaching quality. At the same time, educators can not only teach through serious games, but also track learners' learning process through the management environment created by games, comprehensively monitor students and evaluate their learning effects [6].

Serious games can contribute to E-learning:

1.In terms of teaching:

1) Serious games can optimize teaching methods, improve students' learning efficiency [6], and effectively help students understand and accept new materials.

Serious games can effectively simulate real scenes and provide students with opportunities to learn knowledge and develop abilities without interference and external pressure [7,8,9]. Simulation is an important concept in serious games. Simulating real-life scenes can also immerse students in the game, help them concentrate, achieve deep learning and learn knowledge more effectively. Developers need to carry out appropriate game design to ensure a good simulation. In terms of game tasks, reality can be better simulated through game types such as role-playing games and adventure games. In terms of technology, virtual reality and 3D modeling are usually used to simulate real scenes. For example, in many serious games on medical education topics, the teaching process relies heavily on virtual environments. These serious games simulate scenes in hospitals such as emergency rooms. Players will play the role of nurses or doctors to diagnose and treat virtual patients. Players can practice in a virtual environment, thereby reducing the risks that may occur in reality. At the same time, players gain practical experience, knowledge and skills. For example, in order to practice CPR, players need to perform correct chest compressions on virtual patients in the game [5].

Serious games have good interactivity and can help complete multi-sensory learning. The interactive technology of serious games can effectively improve the player's attention, and can also effectively guide the player to complete game tasks. As introduced in [10], serious games can demonstrate to the player the actions that should be performed through a series of animations. Multi-sensory learning is to mobilize multiple senses to complete learning tasks, usually referring to audio, touch, movement and vision. Studies have proved that children will give priority to using the senses to learn specific concepts, which proves the importance of multi-sensory education [11]. In the learning process, mobilizing students' multiple senses can effectively improve the teaching effect. Multi-sensory learning is usually achieved with the help of sensors and devices.

2) Serious games can improve the teaching process.

Serious games can guide students. Serious games can integrate teaching content into different levels according to the degree of difficulty, and provide a step-by-step, easy-to-understand learning process. At the same time, when the player encounters difficulties during the game, the prompt mechanism of the serious game can guide the player to continue the game.

Serious games can record, monitor, and evaluate the learning process of students. Serious games are a good medium for recording students' learning process and evaluating their learning effects [12]. The database can record the player's relevant information, synchronize the player's game process, and accurately analyze the learning effect according to the player's behavior, such as what learning content the player has achieved, and what knowledge points the player has not fully mastered.

2. In terms of learning:

1) Serious games can improve students' learning motivation.

Serious games will attract players to participate in the game with their fun. The fun of serious games is usually provided by game plot, gameplay and game scene settings. The fun of serious games makes learning content interesting, generates in players an attachment effect and create in students an interest in learning through games [2], thereby attracting players to continue learning tasks and improve their learning motivation.

Serious games can also stimulate players to increase their motivation to learn. For example, serious games can set the appropriate level of difficulty, corresponding rewards and punishments, competition items, and prompts that appear when players encounter problems in the game. These all can play a role in motivating players and can effectively improve their motivation for learning.

2) Serious games can increase players' participation.

For students with learning difficulties and inattention, serious play can effectively ensure that students focus on learning by increasing participation [3]. Participation is usually increased through the interactive features of serious games. The interactive features of serious games create in players an attachment effect [2], thereby attracting players to continue the game.

## 2.2 Game Design of Serious Games

This paper divides the elements of game design into three parts: game scenarios, game mechanics and game technologies (As shown in Table 1).

| Game scenarios | Game screen |
| --- | --- |
| | Game plot |
| | Game tasks |
| Game mechanics | Prompt mechanism |
| | Reward and punishment mechanism |
| | Competition mechanism |
| | Community communication mechanism |
| | Evaluation and scoring mechanism |
| | Data collection mechanism |
| | Mission and level mechanism |
| Game technologies | Sensors |
| | Interactive technologies |
| | The technology of simulating reality scenes |

Table 1 The elements of game design

1. Game scenarios

1) Game screen. In serious games, rich and interesting game graphics can attract players and increase their interest in learning.

2) Game plot. Game plot not only attracts players, but also guides players to perform game tasks.

3) Game tasks. Serious games usually include multiple game tasks. Educational goals are divided into these different game tasks, and each game task corresponds to a specific educational goal.

2. Game mechanics

1) Prompt mechanism. Prompt mechanism can guide players to understand the gameplay. Or, when players encounter a problem during the game, the game will give appropriate prompts to continue the game.

2) Reward and punishment mechanism. The purpose of establishing a reward and punishment mechanism is to make players feel satisfied during the game, so as to continue the game tasks and complete the learning goals.

3) Competition mechanism. The game mechanism includes three types: competition with your own history, competition with virtual characters, and competition with other players. A good competition mechanism can motivate players to complete their learning goals faster and more effectively.

4) Community communication mechanism. The setting of community provides

opportunities for discussion between teachers and students and among students, and helps to improve the sense of interaction in serious games. Students can share and exchange learning experiences with each other in the community.

5) Evaluation and scoring mechanism. Serious games need to evaluate the player's game process to help students analyze their own omissions and deficiencies in knowledge. Check the learner's mastery of learning materials [13]. The assessment should be accurate to the details of the learning process, because if the results of the assessment are more detailed, it will be more helpful to teaching.

6) Data collection mechanism. The data collection is to record the learning process of players so that students can make horizontal comparisons and have a clearer understanding of their learning progress. At the same time, it also provides feedback for educators about the learning processing. Serious games can build a rich database for educators to analyze what knowledge students lack and in which areas students have made progress [6].

7) Mission and level mechanism. Serious games need to further divide the learning content into different tasks and levels, gradually increase the difficulty of the game task, and continuously improve the complexity of the learning content, gameplay and interaction [10]. For different educational content, serious games need to set different tasks and play methods [14]. The result of the teaching activity depends on the player's initial level. Each player's background is different. Therefore, serious games need to consider the initial state of different students and set different levels of difficulty, so that players can choose the difficulty level that suits their initial level and continue learning. As in [14], players can avoid certain tasks in the game and learn only for their own weaknesses.

3. Game technologies

1) Sensors, whose function is mainly to realize system interaction and multi-sensory learning. The sensor can accurately read physical and psychological signals of the players.

2) Common interactive technologies, such as Bluetooth and Wi-Fi. When a serious game involves multiplayer collaborative game tasks, If two players are close, they can be connected via Bluetooth; if father, they need to be connected via Wi-Fi.

3) The technology of simulating reality scenes mainly uses virtual reality technology. The virtual environment generated by digital and computer can provide psychological and sensory immersion [15].
First, evidence shows that virtual reality is an efficient tool to enhance the immersion and interactivity of games [16], and it is also an effective empathy machine for education [17]. Secondly, virtual reality not only helps to achieve immersive learning, but also can materialize abstract knowledge, thereby effectively improving the efficiency and quality of the teaching process, and obtaining good learning results [18]. Finally, virtual reality technology can also solve cognitive and emotional learning problems in a cost-effective manner. The software platform and hardware facilities for realizing virtual reality are also constantly improving. There is also growing interest in adding virtual reality technology to serious games [3]. Common virtual reality devices include: VR glasses, VR headsets, etc. At present, the Desktop VR environment has been effectively used in education for a long time, and Escape Rooms constructed for the realization of virtual reality technology have also been applied to serious games [19]}. Virtual reality technology has broad application prospects in education, training, E-learning and other fields [19].

## 2.3 Educational Goals of Serious Games

The point that distinguishes serious games from ordinary entertaining games is that their main purpose is to teach knowledge and skills. However, in some research on serious games, some problems are often found: the development and design of serious games lack the connection with educators [3]. The teaching of knowledge and skills is an important part and is the goal of serious games. Therefore, serious games should be designed according to educational principles [20]. Therefore, the design of serious games should give full consideration to the educational content and educational goals that they want to convey. In order to help educators design teaching processes and game developers design serious games, it is necessary to use a common method to classify educational content and educational goals. This paper will introduce and use Bloom's taxonomy (the original and revised versions).

In the 1940s, some scholars studied the knowledge of human cognition and information processing and found that the brain uses different mental processes in the process of processing information. Therefore, many educational psychologists tried to apply the results of research to the field of education, and help educators design the teaching process, guide students to be good at mobilizing various cognitive processes, and improve the effect and quality of teaching [21]. And in the 1990s, some people tried to decompose the various fields involved in human learning, resulting in a series of classification methods [22]. Among them, the most famous classification theory of educational goals was put forward by educational psychologist Benjamin Samuel Bloom. Benjamin Samuel Bloom is one of the most influential figures in the field of educational psychology. In his book [23] which has been translated into 22 languages [24], he classified the cognitive skills used in the process of human learning, and this classification method is called Bloom's Taxonomy [25].

This classification system has had a profound impact on the education field [26]. Although it is not the only classification theory of educational goals in educational psychology, it is the most influential. Bloom's Taxonomy is a hierarchical structure [25], which divides educational goals into six parts according to the cognitive process: knowledge, understanding, application, analysis, synthesis and evaluation. As shown in Figure 1, it is a pyramid-shaped structure. Bloom's taxonomy (hereinafter referred to as Bloom's original taxonomy) creates a common language among educators. It distinguishes the level of cognitive skills and emphasizes that only higher levels of cognitive skills can achieve better educational goals and effective teaching effects, and complete the transfer of knowledge and skills [26].

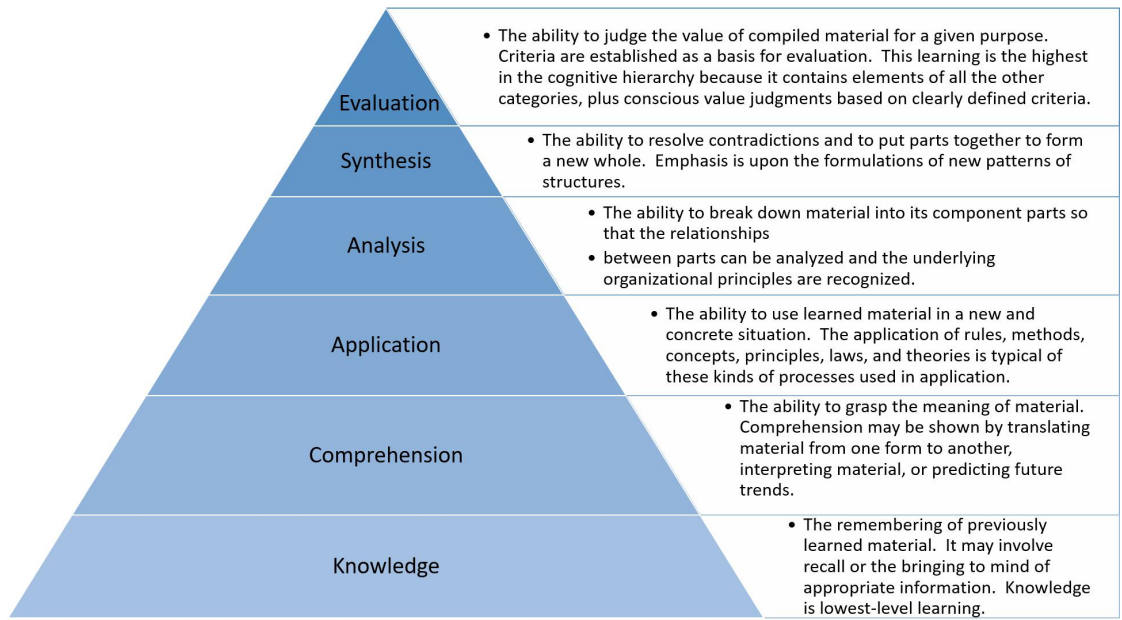

Figure 1 The original Bloom's Taxonomy( [27])

After the original version of Bloom's Taxonomy, Anderson and Krathwohl proposed a revised version based on the original theory in 2001 (hereinafter referred to as the revised Bloom's Taxonomy). It's necessary to revise the original version.

The following changes were made in the revised version: the cognitive process is divided into: Remember; Understand; Apply; Analyze; Evaluate; Create(as shown in Figure 2), and knowledge dimension has been newly added (as shown in Table 2). This dimension defines the types of knowledge: factual knowledge, conceptual knowledge, procedural knowledge and metacognitive knowledge. As shown in Table 2, it has a two-dimensional table structure. In the revised Bloom's Taxonomy, the knowledge dimension helps educators distinguish what type of knowledge is being taught, and the cognitive process dimension helps students retain and transfer the acquired knowledge.

| Knowledge Categories | Definition |
| --- | --- |
| Factual Knowledge | Factual Knowledge is scattered basic knowledge. You just know such a concept and can not really understand it. |
| Conceptual Knowledge | Conceptual Knowledge is a set of interrelated knowledge, and the more relevant knowledge points, the more logical the structure is, and the easier it is to understand and apply. |
| Procedural Knowledge | Procedural Knowledge refers to the steps and methods of doing tasks. |
| Metacognitive Knowledge | Metacognitive Knowledge refers to cognitive and self-cognitive knowledge. |

Table 2 Knowledge categories of the revised version( [22])

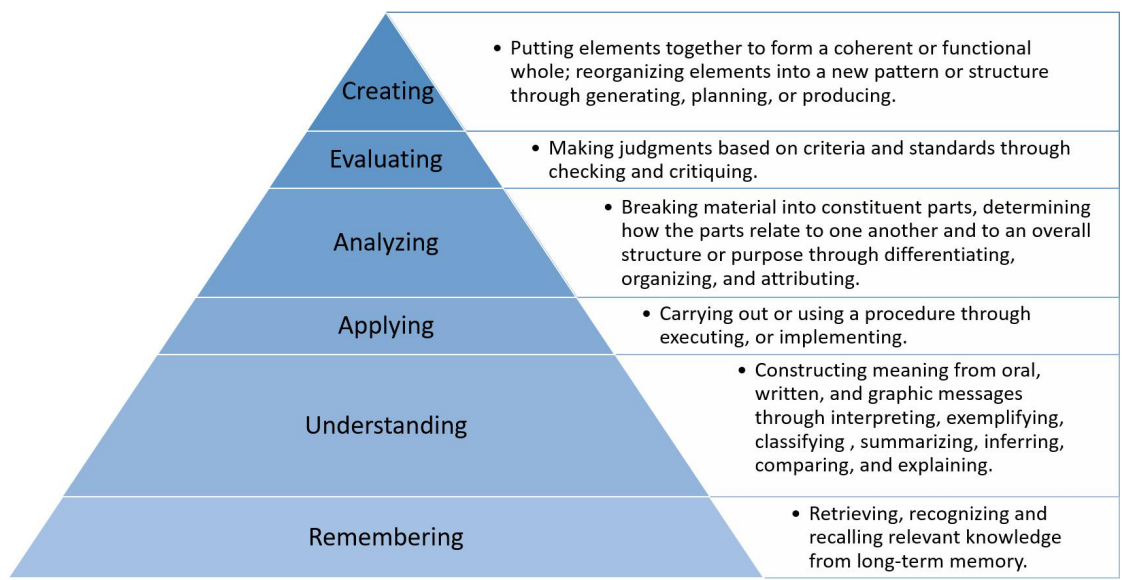

Figure 2 The revised Bloom's Taxonomy( [24])

The revised version provides a more accurate classification and measurement method for the teaching process and the design of serious games. And the classification of these two dimensions is from simple to complex, from easy to difficult. There is a gradual relationship between adjacent categories, with clear levels. The combination of these two dimensions can generate 24 modules, covering all educational goals in the teaching process (as shown in Table 3).

| Knowledge | Cognitive Process | | | | | |
|---|---|---|---|---|---|---|
| | Remembering | Understanding | Applying | Analyzing | Evaluating | Creating |
| Factual Knowledge | | | | | | |
| Conceptual Knowledge | | | | | | |
| Procedural Knowledge | | | | | | |
| Metacognitive Knowledge | | | | | | |

Table 3 The revised version of Bloom's Taxonomy

Facts show that Bloom's classification can be applied to many situations and has universal applicability. The revised Bloom classification is about the common language of learning objectives and a consistent method of teaching design in different countries and regions [28]. The revised classification method provides a concise and clear classification method for the design of the teaching process. Therefore, it can help educators and even game developers analyze and decompose different educational goals, thereby greatly reducing workload.

# 3 The combination of educational goals and game design

According to the above role of game design for serious games and the introduction of various elements of game design, serious games can be classified as following types:

1. Serious games that only motivate players to learn from the game mechanics, such as time limits, competition mechanisms, scoring mechanisms, etc.

The serious game itself does not play a role in optimizing teaching methods. The typical games are simple text games and quiz games.

2. Serious games add fun and playability on the basis of the mechanics of motivation.

In some serious action and puzzle games that are popular with players, game tasks and gameplay can effectively attract players and stimulate their motivation to learn. This type simply combines learning content with games, such as drag and drop games, shooting games, etc.

This type of game is relatively easy to design, and only needs to combine learning content with fun. Therefore, in [29], the concept of modifying existing games into serious games was proposed to teach specific educational goals. That is, simple modifications to existing games such as mechanisms and scenarios can avoid technical obstacles in game development and focus on the delivery of more meaningful educational content.

3. Serious games combine educational content and game design more organically.

This type of serious game will choose game tasks that have natural similarities with the educational goals, and can effectively optimize the teaching methods, and the game process can help players understand and master knowledge. Moreover, such games usually require the use of virtual reality and multi-sensory technology to achieve, and the game scenes of serious games will gradually guide players to complete the prescribed goals.

The elements of the game design and the types of serious games summarized above have some natural connections and compatibility with the four categories of knowledge dimensions of educational goals. Different types of knowledge have different emphasis in the teaching process. Therefore, in corresponding serious games, the elements used in game design are also different.

Factual knowledge is independent, non-systematic knowledge (such as letters, words, symbols, etc.). This type of knowledge focuses on mastering the most basic skills and is relatively simple. Therefore, when students learn factual knowledge, they only need to mobilize the two cognitive processes of remembering and understanding, which can be combined with the first or second type of game. Since factual Knowledge is biased towards the most basic skills, the topics are usually taught in pre-school education and primary education.

Conceptual knowledge is systematic knowledge, such as principles and theories. Compared with factual knowledge, learning this type of knowledge usually requires more complex cognitive processes, which generally include remembering, understanding and applying. It can be used in combination with the second or third game types to enhance the fun of learning and improve teaching methods.

Procedural knowledge refers to knowledge related to the steps and procedures of doing things. Obtaining procedural knowledge requires a lot of practical experience, and the cognitive process of applying is usually used. It can be combined with the third game type, and use simulation (from game scenes and technology) to restore the real scene.

Metacognitive knowledge, that is, knowledge about cognition. At present, there are relatively few serious games on this topic. Acquiring such knowledge usually requires simulation of related scenes to obtain an immersive experience, which can be combined with the third game type.

# 4 Outlook and Discussion

Now, the application of serious games in E-learning also has some problems. An important issue is how to evaluate their effectiveness. Generally, the development and design of serious games need to consume many resources. Thus, it's necessary to ensure that serious games have a stable effect and can be applied to the target group of students. Thus, it is necessary to evaluate and improve the serious game. This paper summarizes the issues that need to be considered and evaluated in the application of serious games in E-learning from several different aspects. First, from the perspective of students, it is necessary to consider whether students are satisfied with serious games, and whether serious games really improve their learning willingness and motivation. For example, players' interest and acceptance of the game scene and game plot, player's initial level, and their ability to adapt to serious games need to be considered. Second, from the perspective of the teaching process, it's necessary to consider whether the theme of game is closely related to the educational goals set by educators, and whether the fun and seriousness of serious games are balanced. In a successful and effective serious game, the game task should be able to promote the completion of the educational goal, and not make the player ignore the learning task because of the gameplay. Third, from the perspective of teaching effect and practical application, it is necessary to consider whether the teaching method of serious games can help students transfer the knowledge and skills learned from the virtual environment into reality, and whether the teaching effect can reach educators' expected goals (that is, compared to traditional E-learning., the teaching method of serious games shows its superiority).

# 6 Conclusion

As a teaching tool, serious games have unique advantages in E-learning. In the development of serious games, the most important issue that needs to be considered is how to balance the serious aspect (that is, the teaching goals and educational content) and the entertaining aspect of serious games. In response to this problem, the main work of this paper is to classify and match the seriousness and entertainment of serious games. And analyze which types of games are suitable for different types of knowledge, so as to provide insight for the development and design of serious games.

games pedagogy," in 2015 IEEE Games Entertainment Media Conference (GEM), 2015, pp. 1–4.

## Authors:

Huansheng Ning received his B.S. degree from Anhui University in 1996 and his Ph.D. degree from Beihang University in 2001. He is currently a Professor and Vice Dean with the School of Computer and Communication Engineering, University of Science and Technology Beijing, China, and the founder and principal at Cybermatics and Cyberspace International Science and Technology Cooperation Base. He has authored 6 books and over 150 papers in journals and at international conferences/workshops. He has been the Associate Editor of IEEE Systems Journal, the associate editor (2014-2018) and the Steering Committee Member of IEEE Internet of Things Journal (2018-), Chairman (2012) and Executive Chairman (2013) of the program committee at the IEEE international Internet of Things conference, and the Co-Executive Chairman of the 2013 International cyber technology conference and the 2015 Smart World Congress. His awards include the IEEE Computer Society Meritorious Service Award and the IEEE Computer Society Golden Core Member Award. His current research interests include Internet of Things, Cyber Physical Social Systems, electromagnetic sensing and computing. In 2018, he was elected as IET Fellow.

Hang Wang received her B.E. degree from Tianjin Normal University and is currently working on her M.S. degree at the School of Computer and Communication Engineering, University of Science and Technology Beijing, China. Her current research focuses on serious games and the Internet of Things.

Wenxi Wang received her B.E. degree from Ludong University in 2019. She is currently pursuing the M.S. degree at the School of Computer and Communication Engineering, University of Science and Technology Beijing. Her current research interests include social computing and artificial intelligence.

Xiaozhen Ye received her B.S. degree from the School of Computer and Communication Engineering, University of Science and Technology Beijing, China, where she is currently pursuing the Ph.D. degree. Her current research interests include serious games and cyber-physical interactions based on video.

Jianguo Ding (jianguo.ding@bth.se) received his degree of a Doctorate in Engineering (Dr.-Ing.) from the faculty of mathematics and computer science at the University of Hagen, Germany. He is currently an Associate Professor at the Department of Computer Science, Blekinge Institute of Technology, Sweden. His research interests include cybersecurity, critical infrastructure protection, intelligent technologies, blockchain, and distributed systems management and control. He is a Senior Member of IEEE (SM'11) and a Senior Member of ACM (SM'20).

Per Backlund holds a PhD from Stockholm University and is currently a

Professor of Information Technology at University of Skovde. He has been an active researcher in the field of serious games since 2005 with a specialization in game based and simulation based training. Professor Backlund has had the role of project manager and principal investigator in several research projects in serious games applications for different application areas such as traffic education, rescue services training and prehospital medicine.